\documentclass{article}
\usepackage{fancyhdr}
\usepackage{diagbox}
\usepackage{multirow}
\usepackage{ragged2e}
\usepackage{amssymb,amsmath}
\usepackage[english]{babel}
\usepackage{fixltx2e}
\usepackage[protrusion=true,expansion=true]{microtype}	
\usepackage{amsmath,amsfonts,amsthm,amssymb}
\usepackage{graphicx}
\graphicspath{ {images/} }
\usepackage{listings}
\usepackage{setspace}
\usepackage{eucal}
\usepackage{array, booktabs}
\usepackage{bm}
\usepackage[justification=centering]{caption}
\usepackage{lipsum}
\usepackage{listings}
\usepackage{appendix}
\usepackage{caption}
\usepackage{subcaption}
\usepackage{bm}
\usepackage{float}
\usepackage{enumitem}
\usepackage{algorithm,algpseudocode}
\usepackage [english]{babel}
\usepackage [autostyle, english = american]{csquotes}
\usepackage[nottoc,notlof,notlot]{tocbibind} 
\addto{\captionsenglish}{}
\usepackage{amsmath,graphicx,mlspconf}
\usepackage [english]{babel}
\usepackage{float}
\usepackage{diagbox}
\usepackage{multirow}
\title{A Fully Convolutional Neural Network Approach \\ to End-to-End Speech Enhancement}
\name{Frank Longueira, Sam Keene}
\address{The Cooper Union\\
franklongueira@gmail.com, keene@cooper.edu}
\begin{document}
%

\maketitle
\begin{abstract}
 This paper will describe a novel approach to the cocktail party problem that relies on a fully convolutional neural network (FCN) architecture.  The FCN takes noisy audio data as input and performs nonlinear, filtering operations to produce clean audio data of the target speech at the output. Our method learns a model for one specific speaker, and is then able to extract that speakers voice from babble background noise.
 
 Results from experimentation indicate the ability to generalize to new speakers and robustness to new noise environments of varying signal-to-noise ratios. A potential application of this method would be for use in hearing aids. A pre-trained model could be quickly fine tuned for an individuals family members and close friends, and deployed onto a hearing aid to assist listeners in noisy environments.
\end{abstract}
\section{Introduction}
One of the largest issues facing hearing impaired individuals in their day-to-day lives is accurately recognizing speech in the presence of background noise \cite{Healy}. While modern hearing aids do a good job of amplifying sound, they do not do enough to increase speech quality and intelligibility. This is not a problem in quiet environments, but a standard hearing aid that simply amplifies audio will fail to provide the user with a signal they can easily understand when the user is in a noisy environment \cite{Dillon}. The problem of speech intelligibility is even more difficult if the background noise is also speech, such as in a bar or restaurant with many patrons. 

While people without hearing impairments usually have no trouble focusing on a single speaker out of multiple, it is a much more difficult task for people with a hearing impairment \cite{Mcdermott}. The problem of picking out one person's speech in an environment with many speakers was dubbed the cocktail party problem \cite{Cherry}. The paper asserts that humans are normally capable of separating multiple speakers and focusing on a single one. However, hearing impaired individuals may have issues when it comes to performing this same task. A solution to the cocktail party problem would be an algorithm that a computer can employ in real-time to enhance the speech corrupted by babble (background noise from other speakers). Traditionally, the cocktail party problem has been approached using several different techniques, such as using microphone arrays, monaural algorithms involving signal processing techniques, and Computational Auditory Scene Analysis (CASA) \cite{Healy}.

Deep learning approaches to the cocktail party problem and speech enhancement in general tend to take noisy spectrograms as input and transform them to clean spectrograms. The use of deep convolutional neural networks and deep denoising autoencoders on spectrograms have proven to be powerful techniques in practice \cite{Simpson}. One drawback to the use of spectrograms as input is the computation of spectrograms tends to be high since the short-time Fourier transform has to be applied to the raw audio data. This prior computation before inputting into the network requires time and hence increases the difficulty of use in real-time applications. In addition, phase information of the input speech tends to be lost in many of these approaches since only the magnitude spectrum is used. This can cause degradation in quality at the output of the system \cite{Paliwal}.

More recent deep learning approaches have considered an end-to-end approach to speech enhancement that requires no feature extraction \cite{Sainath} - \cite{Thickstun}. The noisy time-domain audio signal is used as the input to a neural network and a filtered time-domain audio signal is obtained at the output. This methodology removes the need for a prior STFT computation and retains phase information at the output. This recent push in the deep learning community towards end-to-end speech enhancement systems is one of the motivations for this paper's approach. The other large motivation comes from two papers dealing with the study of CNNs on raw audio data. In the first paper \cite{Fu}, the authors make a strong case for the lack of need for fully connected layers in a neural network that processes raw audio data at the input. Instead, they recommend the use of convolutional layers in order to maintain local correlations in the signal as it passes through the network. In addition, a fully convolutional network (i.e. a CNN with no fully connected layers) will generally have much fewer parameters than a correspondingly similar network that includes fully connected layers. This reduced model complexity is especially important for real-time application of the speech enhancement algorithm. In the second paper \cite{Gong}, the authors provide insight into the inner workings of convolutional layers applied to raw audio data. They make a strong case for the lack of need for pooling layers and emphasize the convolution theorem:
\begin{equation}\label{eq49}
\begin{aligned}
x * h = \mathcal{F}^{-1} \{ \mathcal{F}\{ x \} \cdot \mathcal{F}\{ h \} \}
\end{aligned}
\end{equation}
where $x$ can be viewed as the input audio signal, $h$ is a learned filter, and $\mathcal{F}$ is the Fourier transform operator.
The convolution theorem allows an FCN to be viewed as a large, nonlinear filter bank. By maintaining the size of the raw audio input vector throughout intermediary computations, each filter's output can be viewed as providing a nonlinear filtered representation of the input vector. As the depth of the FCN increases, a larger number of nonlinear filtered representations is achieved. At the final filtering layer, these representations are combined in a matter that rids the input signal of the background noise representations and only keeps the target speech representations. With the motivation of the approach in mind, the next section will provide specific details of the system design.

\section{System Design}
\label{sec:format}

The first step in designing the speech enhancement system is gathering data for training and validation purposes. An openly available audiobook (narrated by a speaker named Pamela) found online serves as the target speech for designing the system \cite{FourGeorges}. In addition, babble noise audio clips were found online to serve as background noise when additively combined to Pamela's speech \cite{BabbleTrain} - \cite{BabbleValidation}. All of these audio clips were downsampled to 16 kHz and to have only one audio channel (taking the element-wise average of the two channels if necessary). Table \ref{table1} concisely describes this data and how it is split for training and validation.
\begin{table}[H]\centering
\resizebox{\columnwidth}{!}{%
\begin{tabular}[t]{lcccc}\toprule
            		&   Target Speech 						&     Babble Noise		& SNR		& Time (Min:Sec)\\ \midrule
Training Set  	&   Chapter 1	&   Bar Noise			&  5 dB		&	35:37			  \\  \addlinespace
Validation Set 	&   Chapter 2	&   Cafe Noise			&  5 dB		& 	5:04				\\ 
\bottomrule
\end{tabular}
}
\caption[Data collection \& splitting]{Data collection and splitting for system design purposes. Target speech refers to Pamela's narration of Chapters 1 - 2 in \cite{FourGeorges}. Babble noise refers to two different environments found online \cite{BabbleTrain} - \cite{BabbleValidation}. Each set of target speech is additively combined with its corresponding set of babble noise at an SNR of 5 dB.} 
\label{table1}
\end{table}

It is important to note that the system is being designed around a single speaker (i.e. Pamela) and a single SNR of 5 dB. The reason for doing this is to first find a few reasonable FCN architectures for the task of denoising Pamela's speech that has been corrupted by babble noise at an SNR of 5 dB. After choosing a subset of FCN architectures, further exploration will be done for denoising Pamela's speech at SNRs of 0 dB and -5 dB in order to choose one FCN architecture for the system. Once a single FCN architecture is selected and fixed, the next step will involve exploring generalization to a new speaker and the system's robustness to different signal-to-noise ratios.

Having decided on training and validation data, designing the system relies on three main things: (1) a methodology for pre-processing the raw audio data for input into the FCN, (2) fixing an FCN architecture, (3) a methodology for post-processing the raw audio data that is output by the FCN. To begin, let's discuss (1). First, target speech is additively combined with its corresponding babble noise. Next, based on stationarity assumptions for speech, the noisy audio data is split into 20 ms frames in which consecutive frames overlap by 50\%.  Each noisy frame is to be multiplied by a corresponding Hanning window of equal length. To complete the pre-processing methodology, for each noisy frame the mean of the entire training target speech is element-wise subtracted and standard deviation of the entire training target speech is element-wise divided. Having fixed the methodology for pre-processing, the methodology for post-processing, (3),  can also be fixed if it is assumed the FCN outputs a preprocessed filtered 20 ms frame of the preprocessed noisy 20 ms input frame. Given an output frame from the FCN, it is element-wise multiplied by the standard deviation of the entire training target speech and the mean of the entire training target speech is added element-wise. Next, the output from the FCN for the next frame (keeping in mind that these two consecutive frames overlap by 50\%) is obtained and the same thing is done. Finally, the overlap-add method of reconstruction is applied to undo the Hanning window that was applied to both overlapping frames. This results in having reconstructed 30 ms of filtered raw audio data that is ready for playback. This post-processing methodology can be iteratively done for an arbitrary amount of noisy input audio data. The most important part of the system design is (2), fixing an FCN architecture.

Since a fully convolutional neural network contains only convolutional layers, the only things to be determined are how deep the network needs to be and the details of each layer (i.e. number of filters, kernel size, etc.). The output of the network is to be a one dimensional vector of the same length as the 20 ms input vector, so two things can be immediately concluded upon. The first conclusion is to use ``same" padding in all layers to ensure the temporal length of the input vector remains the same throughout intermediary computations and therefore at the output. This allows the FCN to be viewed as a nonlinear, filter bank. The second conclusion is the output layer will be a convolutional layer with one filter and no activation function. This output layer is suitable for reconstructing audio data from its nonlinear representations (i.e. the output layer is able to match the range in which audio data exists) and allows for an output that is one-dimensional. Next, the kernel size for all filters in the network will initially be fixed at 5 ms in length, i.e. 25\% of the input size. This can be tuned later on via the validation set. In addition, a dilation factor will not be used in any convolutional layer in order to better preserve local correlations. The structure of each hidden layer will be the following: convolution operation, batch normalization, and ReLU (or PReLU) activation. Batch normalization between the convolution operation and activation function tends to improve training time and generalization performance \cite{Batchnorm}. Also, the ReLU (or PReLU) activation tends to work well in general CNN practice \cite{Goodfellow}. With all of this covered, the only thing left to determine is how many hidden layers are necessary and how many filters per hidden layer. These hyperparameters will be determined by training different FCN architectures and comparing MSE performance on the validation set. All models to follow are supervised trained with Adam SGD to minimize MSE \cite{Adam}. In addition, all training employs early stopping that terminates after 20 epochs of no improvement and returns the parameters of the model with best validation loss during training. 

To begin, an architecture with one hidden layer is trained to find the number of filters needed in a layer. The number of filters is slowly increased and with each new number of filters a model is trained and validation loss computed. This process provides an understanding of how much complexity, in terms of number of filters per layer, is needed for this task.

\begin{table}[H]\centering
\begin{tabular}[t]{ccc}\toprule
Number of Filters   		& Training MSE	& Validation MSE	\\ 	\midrule
	50					&  0.0401				&	0.0541		  	\\  	\addlinespace
	100 					&  0.0384				& 	0.0512			\\ 	\addlinespace
	200 					&  0.0398				&	0.0519			\\ 	\addlinespace
	300 					&  0.0383				&	0.0504		  	\\  	\addlinespace
	400 					&  0.0372				& 	0.0496			\\ 	\addlinespace
	500 					&  0.0377				&	0.0490			\\ 	\addlinespace
	600 					&  0.0381				&	0.0504		  	\\  	\addlinespace
	700 					&  0.0384				& 	0.0503			\\ 	\addlinespace
	800 					&  0.0375				&	0.0507			\\ 	\addlinespace
	900 					&  0.0391				&	0.0497			\\ 	\addlinespace
	1000					&  0.0373				&	0.0498			\\ 	

\bottomrule
\end{tabular}
\caption[Relationship between number of filters and validation loss]{This table describes training loss \& validation loss (MSE) of single hidden layer FCN architectures as the number of filters increases.  }
\label{table2}
\end{table}
Table \ref{table2} shows that increasing the number of filters marginally helps reduce training loss and validation loss. With this result, a similar procedure will be followed to gain an understanding of how the depth of the network improves performance. The procedure involves first fixing the number of filters to be either 50, 100, or 200 filters per layer and then the depth of the network is increased.
\begin{figure}[H]
  \centering
  \includegraphics[width = 8cm, height = 6cm]{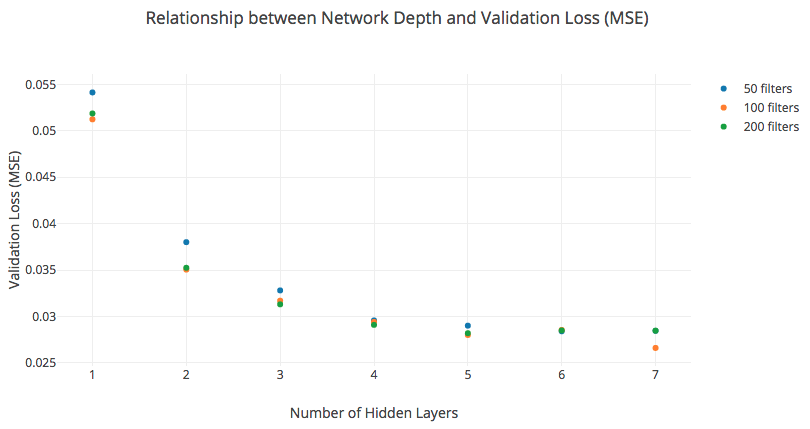}
  \caption[Relationship between network depth and  validation loss]{A plot that shows increasing network depth leads to large decreases in validation loss. Each curve was generated by fixing the number of filters per hidden layer (either 50, 100, or 200) and increasing the number of hidden layers in the network.}
\end{figure}
From Figure 1, it can be concluded that increasing depth largely helps decrease validation loss as compared to increasing the number of filters in a given layer. With the knowledge that between 1 to 200 filters in a layer and a depth of 5 to 6 layers tends to provide good validation loss, different network architectures are experimented with within this range. Architectures with different kernel sizes, activation functions, number of filters and regularization were experimented with with the goal of minimizing validation loss. Over 70 FCN architectures were trained and validated in total. From the over 70 FCN architectures, the top 13 FCN architectures that provide the lowest validation loss are taken to be studied further. The PESQ and WER of the filtered validation set is computed for each of the 13 architectures and compared to the PESQ and WER of the noisy validation set (at an SNR of 5 dB) \cite{PESQ} - \cite{WER}. These results are displayed in Table 3. 

Table 4 presents the architecture from Table 3 (Model \#53) that provided the best combination of high validation PESQ (good speech quality), low validation WER (good intelligibility), and sufficient model complexity for learning across different SNRs. This FCN architecture will be used for testing the speech enhancement system. The next section will present results related to testing the system on new audio data in order to measure its ability to generalize on the same speaker.
\begin{table}[H]\centering
\resizebox{\columnwidth}{!}{%
\begin{tabular}[t]{cccc}\toprule
Model \#  					& Number of Parameters	& 	PESQ 		&	WER\\ 	\midrule
	25					&  3,218,501			&	2.470		&	29.001\% 	\\  	\addlinespace
	26 					&  12,837,001			& 	2.445		&	28.591\%	\\ 	\addlinespace
	27 					&  1,009,501			&	2.390		&	26.402\%	\\ 	\addlinespace
	30 					&  1,209,751			&	2.441		& 	31.464\%	\\  	\addlinespace
	31 					&  4,819,501			& 	2.490		&	31.737\%	\\ 	\addlinespace
	33 					&  2,142,896			&	2.421		&	28.728\%	\\ 	\addlinespace
	38 					&  5,867,728			&	2.452		& 	29.001\%	\\  	\addlinespace
	41 					&  4,682,896			& 	2.422		&	28.454\%	\\ 	\addlinespace
	53 					&  2,266,736			&	2.458		&	25.718\%	\\ 	\addlinespace
	64 					&  761,251			&	2.437		&	29.275\%	\\ 	\addlinespace
	69					&  761,501			&	2.443		&	27.633\%	\\ 	\addlinespace
	70					&  1,562,101			&	2.451		&	29.412\%	\\ 	\addlinespace
	71					&  841,251			&	2.480		&	27.223\%	\\ 	

\bottomrule
\end{tabular}
}
\caption[PESQ \& WER for top 13 FCN architectures based on validation loss]{Number of parameters, PESQ, and WER of top 13 FCN architectures in terms of validation loss. The specific details of each architecture are removed for brevity. For comparison, the noisy validation set at an SNR of 5 dB has a PESQ of 1.764 and WER of 50.479\%.    }
\label{table3}
\end{table}

\begin{table}[!ht]
\centering
\resizebox{\columnwidth}{!}{%

\begin{tabular}[!ht]{cccc}\toprule
Layer Type  						& 	Output Shape		& 	Number of Parameters 	\\ 	\midrule
\hline
1-D Convolution 					&  	(320, 12)		&	972					\\ 	\addlinespace
\hline
Batch Normalization					&  	(320, 12)		&	48					\\ 	\addlinespace
\hline
PReLU Activation 					&  	(320, 12)		&	3,840				\\ 	\addlinespace
\hline
1-D Convolution 					&  	(320, 25)		&	24,025				\\ 	\addlinespace
\hline
Batch Normalization					&  	(320, 25)		&	100					\\ 	\addlinespace
\hline
PReLU Activation 					&  	(320, 25)		&	8,000				\\ 	\addlinespace
\hline
1-D Convolution 					&  	(320, 50)		&	100,050				\\ 	\addlinespace
\hline
Batch Normalization					&  	(320, 50)		&	200					\\ 	\addlinespace
\hline
PReLU Activation 					&  	(320, 50)		&	16,000				\\ 	\addlinespace
\hline
1-D Convolution 					&  	(320, 100)		&	400,100				\\ 	\addlinespace
\hline
Batch Normalization					&  	(320, 100)		&	400					\\ 	\addlinespace
\hline
PReLU Activation 					&  	(320, 100)		&	32,000				\\ 	\addlinespace
\hline
1-D Convolution 					&  	(320, 200)		&	1,600,200				\\ 	\addlinespace
\hline
Batch Normalization					&  	(320, 200)		&	800					\\ 	\addlinespace
\hline
PReLU Activation 					&  	(320, 200)		&	64,000				\\ 	\addlinespace
\hline
1-D Convolution 					&  	(320, 1)		&	16,001				\\ 	

\bottomrule
\end{tabular}
}
\caption[A layer-by-layer description of the selected model's FCN architecture]{A layer-by-layer description of the speech enhancement system's FCN architecture (Model \#53 from Table 3).}
\label{table4b}
\end{table}
\section{Testing Generalization on the Same Speaker}
The first part of testing involves training the speech enhancement system on a speaker and testing it on the same speaker, but the speech will have never been seen by the system nor the babble environment. First, the speech enhancement system trained and validated from the previous section, i.e. using the model from Table \ref{table4b} as the FCN architecture, is used for testing. Five minutes from Chapter 3 from \cite{FourGeorges} will be used as the test target speech and five minutes of a new babble environment  \cite{BabbleTest} will be used for the test background noise. To make sure the testing procedure is clear, consider the following walkthrough of the process. 

First, using the training setup of the previous section, the speech enhancement system is trained using Chapter 1 from \cite{FourGeorges} and bar babble noise from \cite{BabbleTrain} at a specific SNR, say 5 dB. The training process employs early stopping that uses 5 minutes from Chapter 2 from \cite{FourGeorges} and 5 minutes of cafe babble noise from \cite{BabbleValidation} at the same SNR. Next, the trained system is tested on 5 minutes from Chapter 3 in \cite{FourGeorges} and 5 minutes of coffee shop babble noise from \cite{BabbleTest} at the same SNR (i.e. 5 dB), but also at 0 dB and -5 dB to get a measure of the system's robustness across SNRs. This process is repeated for an SNR of 0 dB and an SNR of -5 dB. Tables \ref{table4} \& \ref{table5} report the results of this process.

\begin{table}[H]
\centering
\resizebox{\columnwidth}{!}{%
\begin{tabular}{l*{3}{c}}
\toprule
& \multicolumn{3}{c}{Testing SNR} \\
		& 	5 dB			&  	0 dB						&		-5 dB					\\ 
Training SNR & 					& 							& \\
\hspace{20pt} 5 dB		&	2.478				&	1.917					&		1.350					\\ 
\cmidrule(lr){1-4}
\hspace{20pt} 0 dB		&	2.530				&	2.100					&		1.461					\\
\cmidrule(lr){1-4}
\hspace{19pt} -5 dB	&	2.379				&	2.060					&		1.482					\\ 
\cmidrule(lr){1-4}
		&	(Noisy: 1.782)		&	(Noisy: 1.444)			&		(Noisy: 1.321)			\\
\bottomrule
\end{tabular}
}
\caption[PESQ of speech enhancement system tested on the same speaker across multiple SNRs]{This table presents PESQ test results across different SNRs for testing on the same speaker. Each row represents the SNR that the speech enhancement system was trained at. Each column represents the SNR of the test set. For reference, the PESQ of the noisy test set is included for each SNR.  }
\label{table4}
\end{table}

\begin{table}[H]
\centering
\resizebox{\columnwidth}{!}{%
\begin{tabular}{l*{3}{c}}
\toprule
& \multicolumn{3}{c}{Testing SNR} \\
		& 	5 dB					&  	0 dB						&		-5 dB					\\ 
Training SNR & 					& 							& \\
\hspace{20pt} 5 dB		&	25.243\%				&	60.888\%					&		94.868\%					\\
\cmidrule(lr){1-4}
\hspace{20pt} 0 dB		&	31.900\%				&	58.391\%					&		91.817\%					\\
\cmidrule(lr){1-4}
\hspace{17pt} -5 dB	&	52.705\%				&	77.531\%					&		94.452\%					\\ 
\cmidrule(lr){1-4}
		&	(Noisy: 43.689\%)		&	(Noisy: 86.269\%)			&		(Noisy: 97.503\%)			\\ 
\bottomrule
\end{tabular}
}
\caption[WER of speech enhancement system tested on the same speaker across multiple SNRs]{This table presents WER test results across different SNRs for testing on the same speaker. Each row represents the SNR the speech enhancement system was trained at. Each column represents the SNR of the test set. For reference, the WER of the noisy test set is included for each SNR.  }
\label{table5}
\end{table}

Tables \ref{table4} \& \ref{table5} provide some interesting insights into the generalizability of the speech enhancement system. The first key insight is that both PESQ and WER on the test set do a good job tracking the results for PESQ and WER seen on the validation set. The other key insight is that the speech enhancement system trained at 0 dB is quite robust across SNRs, sometimes doing better in scenarios one would not expect. With promising results from testing on the same speaker, the second part of testing will study generalizability to a new speaker.
\section{Testing Generalization on a New Speaker}
The second part of testing will employ the same methodology as the first part of testing but the target speech will come from a new speaker. Audio data is acquired for a new speaker (specifically another female speaker by the name of Tricia) via another audiobook \cite{Twain}. First, performance will be measured when the speech enhancement system is trained only on the speaker Pamela (as has been done up to this point) and tested on the speaker Tricia. All trained models (i.e. at each specific SNR) from the first part of testing are used again in the second part of testing. These trained models filter 5 minutes of speech from Chapter 1 of \cite{Twain} corrupted by the same coffee shop babble noise \cite{BabbleTest} at SNRs of -5, 0, and 5 dB. Tables \ref{table6} \& \ref{table7} report the results of this process.

\begin{table}[H]
\centering
\resizebox{\columnwidth}{!}{%
\begin{tabular}{l*{3}{c}}
\toprule
& \multicolumn{3}{c}{Testing SNR} \\
		& 	5 dB						&  	0 dB						&		-5 dB				\\ 
Training SNR & 					& 							& \\
\hspace{20pt} 5 dB		&	2.215					&		1.781				&		1.291					\\ 
\cmidrule(lr){1-4}
\hspace{20pt} 0 dB		&	2.171					&		1.839				&		1.382					\\ 
\cmidrule(lr){1-4}
\hspace{19pt} -5 dB	&	2.229					&		1.876				&		1.437				\\ 
\cmidrule(lr){1-4}
		&	(Noisy: 1.874)			&	(Noisy: 1.471)			&		(Noisy: 1.182)			\\
\bottomrule
\end{tabular}
}
\caption[PESQ of speech enhancement system trained on one speaker and tested on a new speaker across multiple SNRs]{This table presents PESQ test results across different SNRs for training on one speaker and testing on a new speaker.}
\label{table6}
\end{table}

\begin{table}[H]
\centering
\resizebox{\columnwidth}{!}{%
\begin{tabular}{l*{3}{c}}
\toprule
& \multicolumn{3}{c}{Testing SNR} \\
		& 	5 dB						&  	0 dB							&		-5 dB					\\ 
Training SNR & 					& 							& \\
\hspace{20pt} 5 dB		&	25.134\%					&	54.545\%						&		89.037\%					\\
\cmidrule(lr){1-4}
\hspace{20pt} 0 dB		&	29.412\%					&	50.401\%						&		84.358\%					\\
\cmidrule(lr){1-4}
\hspace{17pt} -5 dB	&	55.615\%					&	67.380\%						&		89.305\%					\\ 
\cmidrule(lr){1-4}
		&	(Noisy: 33.556\%)		&	(Noisy: 72.326\%)			&		(Noisy: 94.652\%)			\\ 
\bottomrule
\end{tabular}
}
\caption[WER of speech enhancement system trained on one speaker and tested on a new speaker across multiple SNRs]{This table presents WER test results across different SNRs for training on one speaker and testing on a new speaker.}
\label{table7}
\end{table}
When comparing Tables \ref{table4} \& \ref{table5} with Tables \ref{table6} \& \ref{table7}, it is noticed that performance on the new speaker is good but does not quite track the performance of a system trained on a speaker and then tested on that same speaker. It is hypothesized that fine-tuning the parameters of the trained network with a few minutes of data from the new speaker should improve performance and more closely track performance on the same speaker. Therefore, an additional, disjoint 5 minutes of speech from Chapter 1 of \cite{Twain} corrupted by 5 minutes of cafe babble noise from \cite{BabbleValidation} at a given SNR is used to fine-tune the already trained model (i.e. trained on the speaker Pamela) by doing 5 epochs of gradient descent. The resulting trained model is then used to again filter 5 minutes of speech from Chapter 1 of \cite{Twain} corrupted by the same coffee shop babble noise \cite{BabbleTest} at SNRs of -5, 0, and 5 dB. Tables \ref{table8} \& \ref{table9} report the results of this process.

\begin{table}[H]
\centering
\resizebox{\columnwidth}{!}{%
\begin{tabular}{l*{3}{c}}
\toprule
& \multicolumn{3}{c}{Testing SNR} \\
		& 	5 dB						&  	0 dB						&		-5 dB				\\ 
Training SNR & 					& 							& \\
\hspace{20pt} 5 dB		&	2.417					&	1.953					&		1.442					\\ 
\cmidrule(lr){1-4}
\hspace{20pt}  0 dB		&	2.378					&	2.025					&		1.571					\\ 
\cmidrule(lr){1-4}
\hspace{19pt}-5 dB	&	2.283 					&	2.026					&		1.619					\\ 
\cmidrule(lr){1-4}
		&	(Noisy: 1.874)			&	(Noisy: 1.471)			&		(Noisy: 1.182)			\\ 
\bottomrule

\end{tabular}
}
\caption[PESQ of speech enhancement system trained on one speaker, fine-tuned on a new speaker, and tested on that new speaker across multiple SNRs]{This table presents PESQ test results across different SNRs for training on one speaker, fine-tuning on a new speaker, and then testing on that new speaker.}
\label{table8}
\end{table}

\begin{table}[H]
\centering
\resizebox{\columnwidth}{!}{%
\begin{tabular}{l*{3}{c}}
\toprule
& \multicolumn{3}{c}{Testing SNR} \\
		& 	5 dB					&  	0 dB						&		-5 dB \\
Training SNR & 					& 							& \\
\hspace{20pt}  5 dB		&	21.791\%			   	&	45.588\%					&	87.166\%			\\
\cmidrule(lr){1-4}
\hspace{20pt} 0 dB		&	28.342\%		      	 	&	40.909\%					&	76.337\%			\\ 
\cmidrule(lr){1-4}
\hspace{17pt}  -5 dB	&	58.690\%			   	&	79.813\%					&	81.684\%			\\
\cmidrule(lr){1-4}
		&	(Noisy: 33.556\%)		&	(Noisy: 72.326\%)			&	(Noisy: 94.652\%)	\\ 
\bottomrule
\end{tabular}
}
\caption[WER of speech enhancement system trained on one speaker, fine-tuned on a new speaker, and tested on that new speaker across multiple SNRs]{This table presents WER test results across different SNRs for training on one speaker, fine-tuning on a new speaker, and then testing on that new speaker.}
\label{table9}
\end{table}

When comparing Tables \ref{table4} \& \ref{table5} with Tables \ref{table8} \& \ref{table9}, it is noticed that performance on the new speaker now does a good job tracking the performance of a system trained on a speaker then tested on that same speaker. This helps verify the initial hypothesis and it can be concluded that the speech enhancement system trained at 0 dB is able to generalize to new speakers (via fine-tuning) and is markedly robust across different SNRs.


\newpage

\section{Conclusions \& Future Work}
A fully convolutional neural network based end-to-end speech enhancement system that serves as a solution to the famous cocktail party problem has been presented. An ability to generalize to new speakers is presented by fine-tuning of the system with limited data. Test results show that the system is robust to different babble noise environments of varying SNRs. This speech enhancement system shows promising results objectively, using PESQ and WER measures, and subjectively by listening to the filtered audio. A few questions to consider for future research pertaining to this speech enhancement system are presented below:
\bigskip

\noindent 1) What is the optimal model complexity for the task? \\
2) Does the system continue to generalize well to new environments and speakers? \\
3) What is the minimal computational/storage complexity needed to employ this system in real-time?

\singlespacing

\bibliographystyle{IEEEbib}
\bibliography{strings,refs}

\end{document}